\documentclass[a4paper,10pt]{article}
\usepackage[utf8x]{inputenc}
\usepackage{amsthm,amssymb,amsmath}
\usepackage{graphicx}
\usepackage{epsfig}

\newcommand{\oh}{\Omega h^2}
\newcommand{\gev}{\mathrm{GeV}}
\newcommand{\kev}{\mathrm{keV}}
\newcommand{\tev}{\mathrm{TeV}}
\newcommand{\sv}{\langle\sigma v\rangle}

\newcommand{\mh}{M_{H_2}}
\newcommand{\mhone}{M_{H_1}}
\newcommand{\sa}{\sin\alpha}
\newcommand{\mdm}{M_\chi}

\newcommand{\tfi}{T_{f.i.}}
\newcommand{\tfo}{T_{f.o.}}

\title{Warm and cold  fermionic dark matter via freeze-in}
\author{Michael Klasen\footnote{michael.klasen@uni-muenster.de}  ~and 
Carlos E. Yaguna\footnote{carlos.yaguna@uni-muenster.de} \\ 
\it \small Institut f\"ur Theoretische Physik, Universit\"at M\"unster,\\
\it \small Wilhelm-Klemm-Stra\ss e 9, D-48149 M\"unster, Germany}
\date{}  

\begin{document}
\maketitle
\vspace*{-8cm}
\begin{flushright}
\texttt{MS-TP-13-24}
\end{flushright}
\vspace*{7cm}
\begin{abstract}
The freeze-in mechanism of dark matter production provides a simple and intriguing alternative to the WIMP paradigm. In this paper, we analyze whether freeze-in can be used to account for the dark matter in the so-called   singlet fermionic model. In it, the SM is extended with only two additional fields, a singlet scalar that mixes with the Higgs boson, and the dark matter particle, a fermion assumed to be odd under a $Z_2$ symmetry. After numerically studying the generation of dark matter, we analyze  the dependence of the relic density with respect to all the free parameters of the model. These results are then used to obtain the regions of the parameter space that are compatible with the dark matter constraint. We demonstrate that the observed dark matter abundance can be explained via freeze-in over a wide range of masses extending down to the keV range. As a result, warm and cold dark matter can be obtained in this model. It is also possible to have dark matter masses well above the unitarity bound  for WIMPs.     
\end{abstract}

\section{Introduction}
Not much is currently known about the fundamental properties of the dark matter. The most common assumption is that it consists of Weakly Interacting Massive Particles (WIMPs), which not only appear in different extensions of the Standard Model (SM) but have also the advantage of explaining in a natural way the observed dark matter density --via the so-called freeze-out process. In the early Universe, WIMPs are initially in thermal equilibrium but as  the temperature decreases below the dark matter mass, it becomes more difficult to maintain their equilibrium distribution and at some temperature, known as the freeze-out temperature $\tfo$, the abundance departs from equilibrium and freezes out, remaining constant until today. It turns out that for a particle with weak-strength interactions and a mass in the hundreds of GeVs (a WIMP), this freeze-out process generally leads to a dark matter density close the observed value \cite{Hinshaw:2012aka,Ade:2013lta}. This fact, sometimes referred to as the WIMP miracle, partially explains why in most models considered in the literature the dark matter particle is a WIMP. The other reason is that WIMP dark matter candidates can be easily found in well-known extensions of the SM. In the MSSM, for example, the neutralino is the dark matter candidate and has become the archetype of WIMP dark matter. In spite of all this,  the WIMP framework for dark matter should no be taken for granted and must instead be regarded as an interesting hypothesis to be tested in current and future experiments. WIMPs  should give rise to a number of signatures in accelerator searches  and in direct and indirect dark matter detection experiments, none of which have been observed --notwithstanding a huge experimental effort on all these fronts.  Even if the current data does not yet rule out the possibility that dark matter is composed of WIMPs, this situation might soon change thanks to the LHC upgrade and to the next generation of dark matter experiments.  In fact, it was recently claimed that the moment of truth for WIMPs had finally arrived \cite{Bertone:2010at}: if they  are not discovered in the next few years the WIMP paradigm will have to be abandoned and some other way of explaining the dark matter will have to be found.

Viable and well-motivated alternatives to the WIMP framework certainly exist \cite{Choi:2005vq,Feng:2003xh,Hall:2009bx}. Among them the so-called \emph{freeze-in} mechanism \cite{Hall:2009bx} looks particularly promising. It involves a Feebly Interacting Massive Particle (FIMP) which never attains thermal equilibrium in the early Universe. FIMPs are slowly produced by collisions or decays of particles in the thermal plasma and, in contrast to WIMPs, are never abundant enough to annihilate among themselves. Thus, the abundance of FIMPs freezes in with a yield that \emph{increases} with the interaction strength between them and the thermal bath. FIMPs, in addition, give rise to   completely different signatures in colliders and dark matter experiments \cite{Hall:2009bx}. Due to its simplicity and predictive power, the \emph{freeze-in} mechanism of dark matter production is a strong contender to the WIMP framework.

Few explicit realizations of the freeze-in mechanism have been studied in the literature \cite{McDonald:2001vt,Asaka:2005cn,Asaka:2006fs,Hall:2009bx,Yaguna:2011qn,Yaguna:2011ei,Chu:2011be,Cheung:2011nn}. In \cite{Hall:2009bx}, for example, different  possibilities were briefly considered, all of them related to known scenarios for physics beyond the SM, such as supersymmetry and extra-dimensions. Another approach is to incorporate this mechanism into minimal extensions of the SM that can also account for the dark matter. Given that a FIMP has to be a singlet under the SM gauge group (to prevent it from reaching thermal equilibrium in the early Universe), the two main  possibilities are a singlet scalar and a singlet fermion.  The  former was already considered in \cite{Yaguna:2011qn,Yaguna:2011ei}, where it was shown that it can indeed explain the dark matter in the FIMP regime. The latter, a singlet fermion, is the subject of this paper. Specifically, we consider the so-called singlet fermionic model of dark matter, in which two fields are added to the SM: the dark matter candidate --a fermion odd under a $Z_2$-- and a singlet scalar that is even under the $Z_2$ and mixes with the SM Higgs boson.  The WIMP regime of this model is well-known as it has been studied in several works --see e.g. \cite{Kim:2008pp, Baek:2011aa, LopezHonorez:2012kv, Baek:2012uj, Fairbairn:2013uta, Esch:2013rta}. In this paper, we will, for the first time, examine the feasibility of accounting for the  dark matter, via the freeze-in mechanism,  in the singlet fermionic model of dark matter.

As we will see, the freeze-in mechanism allows to explain the observed relic density over a wide range of dark matter masses, from the  $\kev$ scale, where the dark matter particle behaves as \emph{warm} dark matter, to the more familiar TeV scale usually associated with WIMPs, and going even further to  super-heavy dark matter, with masses exceeding the weak scale by several orders of magnitude.  Thus, both cold and warm dark matter can be obtained in this model. The issue of whether warm dark matter is actually needed by or consistent with current data remains unsettled (see e.g. \cite{Biermann:2013nxa,Viel:2013fqw,Lovell:2011rd}), and we have nothing to add to that discussion.  But if it is, it is important to keep in mind that the singlet fermion we consider offers an interesting alternative to the sterile neutrino \cite{Canetti:2012kh} and has the advantage, by virtue of the $Z_2$ symmetry, of being absolutely stable, evading the strong bounds from X-ray observations to which the sterile neutrino is subject \cite{Watson:2011dw}.

The rest of the paper is organized as follows. In the next section we introduce our notation, present the model  and discuss its free parameters.  Then in section \ref{sec:fi} we briefly review the freeze-in mechanism for dark matter production, with a special emphasis on the singlet fermionic model. Section \ref{sec:res} includes our new results. In it we first study the evolution of the  dark matter abundance and its dependence with the free parameters of the model. Then, we analyze the prediction for the relic density of dark matter in this model. Next, the viable regions of the freeze-in realization of the singlet fermionic model are presented. Finally, we briefly discuss some implications of our results in section \ref{sec:dis} and draw our conclusions in section \ref{sec:con}. 

\section{The model}
\label{sec:mod}
The model we consider is a minimal extension of the SM with two additional fields, one Majorana fermion, $\chi$, and one real scalar, $\phi$. $\chi$, the dark matter candidate,  is a singlet under the SM gauge group and odd under a $Z_2$ symmetry that guarantees its stability --all other particles including $\phi$ are even under the $Z_2$.   The only renormalizable interactions that the dark matter particle can have in this model are Yukawa terms of the form  $\bar \chi\chi\phi$. Since $\phi$ is even under the $Z_2$, it will mix with the Higgs boson providing a link between the dark matter sector and the SM.

Besides  the kinetic term, the part of the Lagrangian involving the dark matter particle $\chi$ is given by   
\begin{equation}
 \mathcal{L}_\chi=-\frac12M_\chi \bar\chi\chi+g_s\phi\bar\chi\chi+ig_p\phi\bar\chi\gamma_5\chi,
\end{equation}
where $\mdm$ is the dark matter mass, $g_s$ is the scalar coupling and $g_p$ is the pseudo-scalar one. For definiteness, we will concentrate on the parity-conserving case and set $g_p=0$ in the following\footnote{We have checked that, in contrast to the WIMP regime, there are no significant differences between the results for $g_p=0$ and $g_p\neq0$.}. In addition, the scalar potential is  modified and now reads
\begin{align}
 V(\phi,H)=&-\mu_H^2H^\dagger H + \lambda_H (H^\dagger H)^2-\frac{\mu_\phi^2}{2} \phi^2+\frac{\lambda_\phi}{4} \phi^4+\frac{\lambda_4}{2}\phi^2H^\dagger H \nonumber\\
& + \mu_1^3\phi+\frac{\mu_3}{3}\phi^3+\mu\, \phi (H^\dagger H),
\label{eq:V}
\end{align}
where $H$ is the usual SM Higgs doublet that breaks the electroweak symmetry after acquiring a vacuum expectation value. In the unitary gauge we have that $ H=\frac {1}{\sqrt 2} \begin{pmatrix} 0\\ v+h\end{pmatrix}$ and $\langle H\rangle=\frac {1}{\sqrt 2} \begin{pmatrix} 0\\ v\end{pmatrix}$. In principle,  $\phi$ can also acquire a VEV, but it is possible to choose a basis (by shifting the field) in such a way that $\langle\phi\rangle=0$. That is the basis in which we  will work throughout this paper. The $\mu$ term in (\ref{eq:V}) induces a mixing between $h$ and $\phi$ which gives rise to two scalar mass eigenstates, $H_1$ and $H_2$, defined as
\begin{equation}
 H_1= h\,\cos\alpha+\phi\,\sin\alpha,\quad H_2= \phi\,\cos\alpha  -h\,\sin\alpha,
\end{equation}
where $\alpha$ is the mixing angle. Notice that  for small mixing $H_1$ becomes a SM-like Higgs, so we will require, in agreement with recent measurements at the LHC \cite{Aad:2012tfa,Chatrchyan:2012ufa}, that $\mhone=125~\gev$ and small $\sa$ --see \cite{Esch:2013rta} for details. The free parameters of this model can be taken to be
\begin{equation}
 \mdm,\mh,g_s,\sin\alpha, \lambda_4, \mu_3.
\end{equation}
Our goal is to find, within this multidimensional parameter space, the regions that are consistent with the dark matter constraint via the freeze-in mechanism. Unless otherwise stated, we assume that $\lambda_4=0$ and $\mu_3=0$ so that our \emph{default} parameter space is four-dimensional and consists only of $\mdm$, $\mh$, $g_s$ and $\sa$. In any case, that assumption will be relaxed in some selected figures so as to illustrate the effect of $\lambda_4$ and $\mu_3$. Before presenting our results, let us briefly review the freeze-in mechanism.

\section{The freeze-in mechanism of dark matter production}
\label{sec:fi}

\begin{figure}[t]
\begin{tabular}{ccc}
 \includegraphics[scale=0.3]{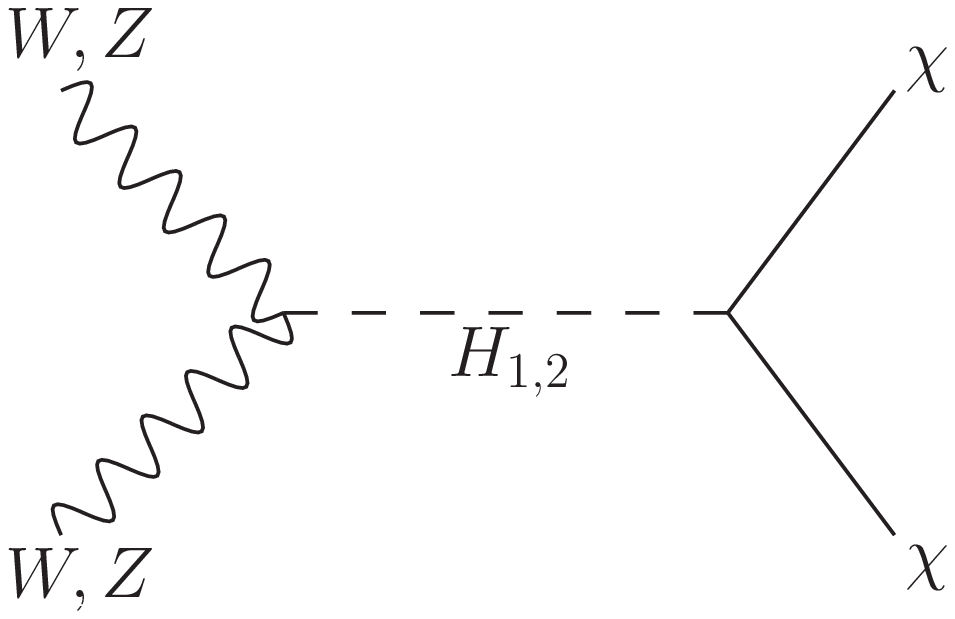} &    \includegraphics[scale=0.3]{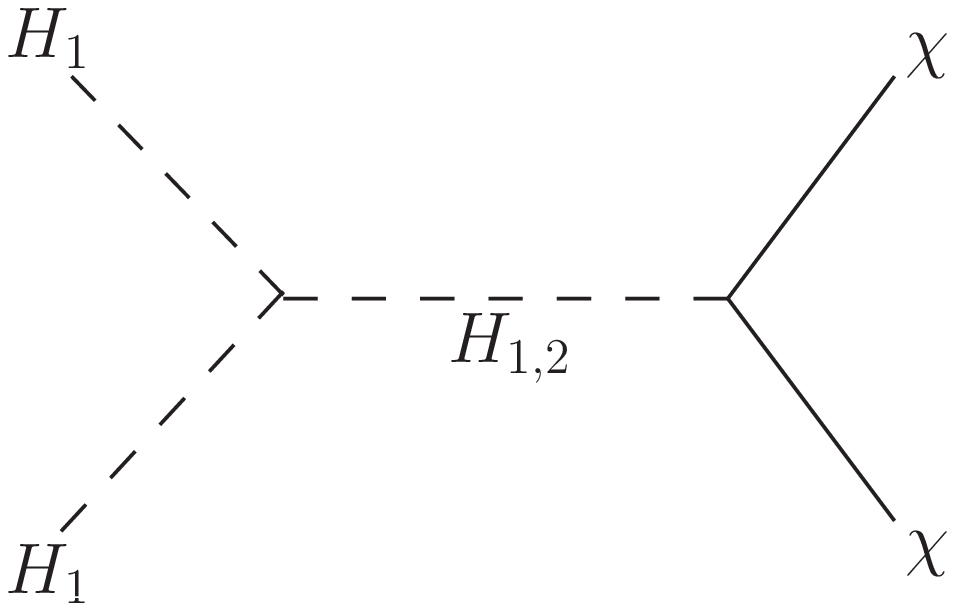} & \includegraphics[scale=0.3]{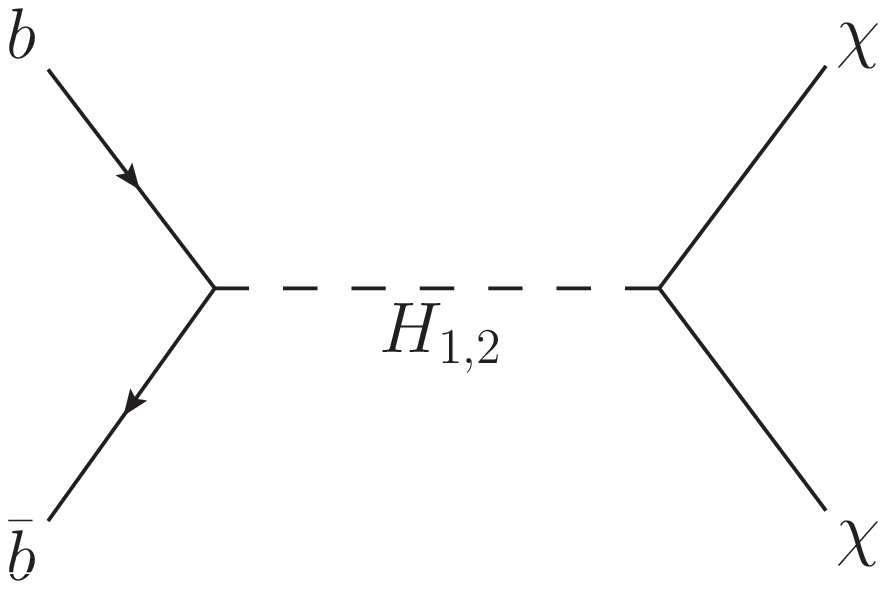} \\
(a) &    (b) & (c)\\
\includegraphics[scale=0.3]{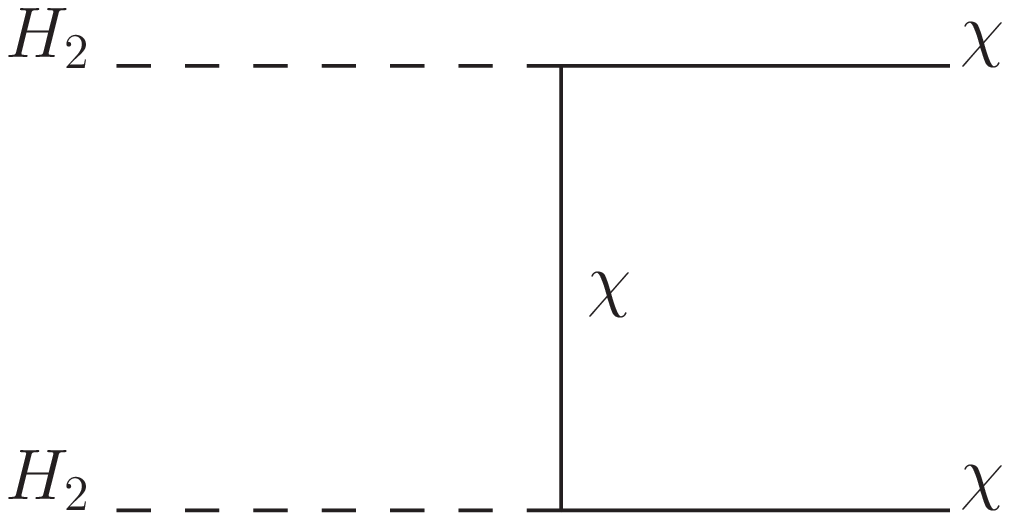}  & \includegraphics[scale=0.3]{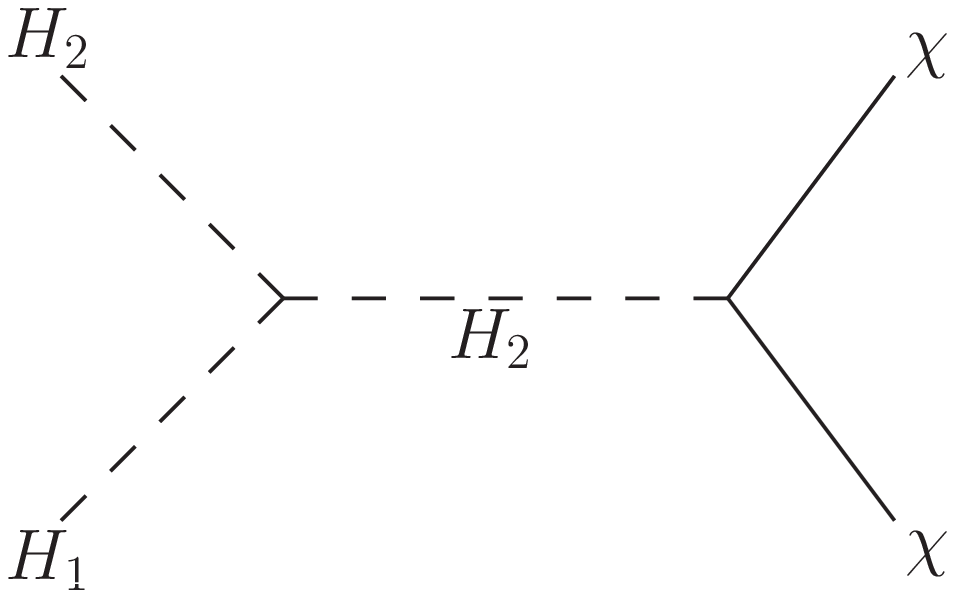} &    \includegraphics[scale=0.3]{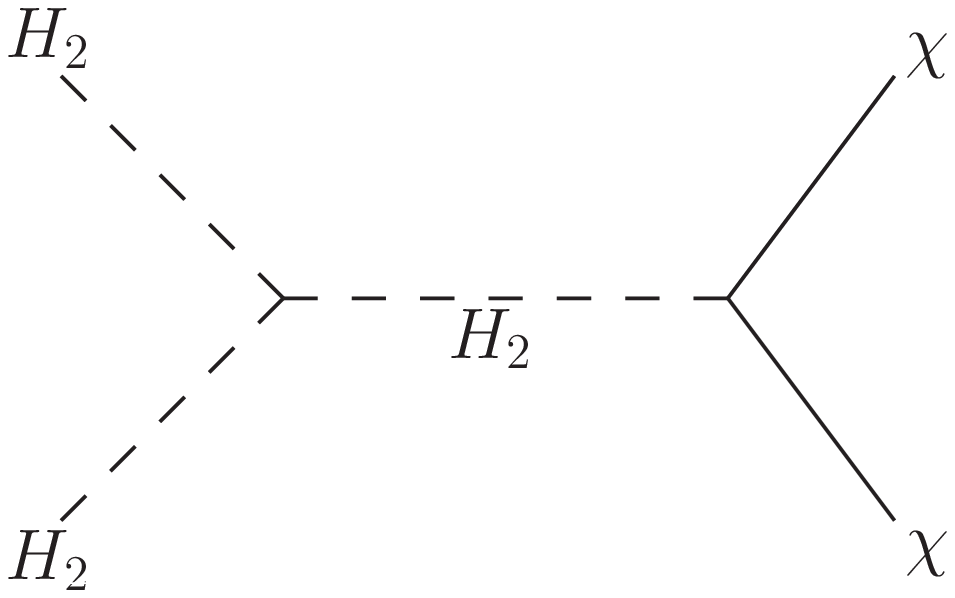} \\
(d) & (e) & (f)
\end{tabular}
\caption{\small Main contributions  to dark matter production in the singlet fermionic model. The dominant contributions over most of the parameter space are  shown in the first row. Such diagrams feature a  cross section that  goes as $g_s^2\sin^2\alpha$ and may be resonantly enhanced.  Diagram (d) gives rise to a cross section that is independent of $\sa$ and proportional to $g_s^4$. Diagrams (e) and (f) become relevant only for $\lambda_4\neq0$ or $\mu_3\neq0$. }
\label{fig:diag1} 
\end{figure}

In the FIMP regime, dark matter particles do not reach equilibrium in the early Universe and are never abundant enough to annihilate among themselves. As a result, the dark matter abundance, $Y=n/s$, is not determined by the same processes as in the WIMP framework. If dark matter particles are pair produced, as it happens in the singlet fermionic model we are studying,  $Y$ satisfies instead the Boltzmann equation
\begin{equation}
\frac{dY}{dT}=-\sqrt{\frac{\pi g_*(T)}{45}}M_p\langle\sigma v\rangle Y_{eq}(T)^2
\label{eq:boltzmann}
\end{equation}
with  the boundary condition $Y(T\gg \mdm)=0$. In the above equation $M_p$ is the Planck mass and  $\langle\sigma v\rangle$ is the usual thermally averaged production (or annihilation) cross section. It is through this quantity that the dependence on the particle physics model enters into the evolution of the dark matter abundance.  The main difference between the WIMP and the FIMP regimes of the singlet fermionic model is the value of the relevant coupling --$g_s$ in the singlet fermionic case--  or equivalently of $\langle\sigma v\rangle$, which, being much smaller for FIMPs, prevents them from ever reaching thermal equilibrium: $Y\ll Y_{eq}$. For that reason, the annihilating term proportional to $Y^2$ does not contribute to the right-hand side of equation (\ref{eq:boltzmann}), and it  cannot be assumed, as is the case for  WIMPs, that the dark matter particle was in equilibrium, $Y(T)=Y_{eq}(T)$, for $T\sim \mdm$.

Using equation (\ref{eq:boltzmann}) we can anticipate the generic behavior of $Y(T)$ in a large class of situations within the singlet fermionic model.  Since the right-hand side is less than zero, as the temperature decreases the abundance either increases or remains constant but never diminishes,  in agreement with the expectation that dark matter annihilations, which would reduce the abundance, play no role in this regime.  Thus, the dark matter abundance initially increases as the Universe cools down but at a certain point it reaches the \emph{freeze-in} temperature, $\tfi$, below which the abundance no longer changes.

Integrating equation (\ref{eq:boltzmann}) from $T_i\gg \mdm$ to $T_f< \tfi$ leads to the  current abundance of dark matter, $Y(T_0)$. From it, the dark matter relic density can be calculated in the usual way:
\begin{equation}
\oh=2.742\times 10^8 \frac{\mdm}{\gev} Y(T_0). 
\label{eq:rd}
\end{equation}

In the singlet fermionic model different processes may contribute to the production of dark matter --see figure \ref{fig:diag1}. In most cases, the dominant processes are the annihilation of $W^+W^-$, $ZZ$, $H_1H_1$ or  $b\bar b$  via $H_1$ and $H_2$ mediated diagrams into $\chi\chi$ --diagrams (a), (b) and (c) in figure \ref{fig:diag1}. The amplitude for each of these diagrams is proportional to $g_s\sin\alpha$. Naively one would expect, based on  dimensional arguments, that at high temperatures   $\sv\propto 1/T^2$, as it happens for the singlet scalar case. That is not necessarily the case in this model, however,  because  the contributions from the $H_1$-mediated diagrams tends to cancel against those mediated by $H_2$,  leaving a  $\sv$ that goes instead as $1/T^4$ over  a wide range of temperatures. Since at high temperatures $Y_{eq}(T)$ is constant,  we expect $Y(T)\propto 1/T^3$ in that region. As the temperature is further increased, that cancellation becomes more precise with the result that, if $g_s$ is not too small, diagram (d), whose amplitude goes like $g_s^2$ and is independent of $\sa$, becomes dominant. If that is the case, the naively expected behaviors, $\sv\propto 1/T^2$ and  $Y(T)\propto 1/T$, are recovered.

Diagrams (a), (b) and (c)  allow, provided that $2\mdm<\mh,\mhone$, for the resonant production of dark matter, an effect of great importance in our results. We will see, in fact, that the predicted relic density and the viable parameter space are very sensitive to the position of the $H_2$ resonance.  It is therefore crucial to correctly take into account the effect of the resonance on $\sv$. To that end, we have implemented the singlet fermionic model into  micrOMEGAs \cite{Belanger:2013oya} (via LanHEP \cite{Semenov:2010qt}) and have used it to calculate $\sv$. This value was then plugged into the Boltzmann equation  for the FIMP regime, equation (\ref{eq:boltzmann}), which we numerically solve. In this way, we guarantee that all processes contributing to dark matter production are taken into consideration (not only those shown in figure \ref{fig:diag1}) and that our results are fairly accurate.

If we allow $\lambda_4$ and $\mu_3$ to be different from zero,  additional diagrams, e.g. (e) and (f) in figure \ref{fig:diag1},  may contribute to dark matter production. Notice that their cross sections are independent of $\sa$, proportional to $g_s^2$, and never resonantly enhanced. As we will show in the next section, they increase the relic density only in selected regions of the parameter space.  

With this background, we turn now to our main results. In the next section we will study the dependence of $Y$ and the relic density on the different parameters of the singlet fermionic model, and the regions consistent with the dark matter constraint will be determined.  
 
\section{Results}
\label{sec:res}

For clarity we have divided our results in three subsections. The first one, section \ref{sec:y}, is dedicated to the study of the dark matter abundance, $Y$. In it we specifically study the variation of $Y$ with the temperature for different sets of  parameters, and we will confirm some of the analytical results obtained in the previous section. Section \ref{sec:rd} concerns  the analysis of the relic density.  The value of the relic density is calculated over a wide range of dark matter masses (from $\kev$ to $10^8~\gev$) for specific values of the remaining parameters. By visually comparing $\oh$ against its measured  value, we get in this section a first impression of the regions that are consistent with the dark matter constraint. In section \ref{sec:vr}, these regions are precisely obtained and projected onto the plane ($\mdm$, $g_s$).

\subsection{The dark matter abundance}
\label{sec:y}

\begin{figure}[t]
\begin{center} 
\includegraphics[scale=0.4]{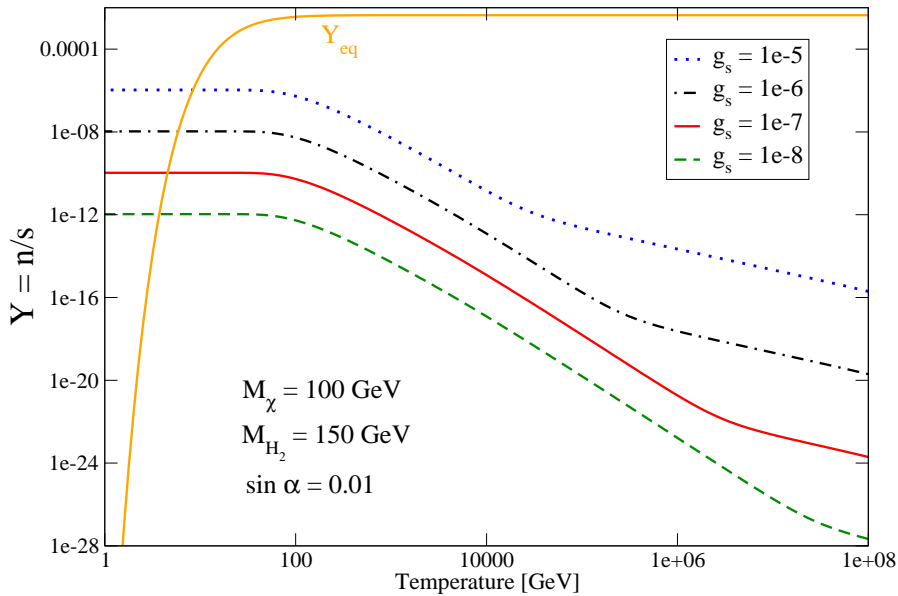}
\caption{\small The dark matter abundance as a function of the temperature for different values of $g_s$. The solid orange line shows $Y_{eq}$. In this figure  $\mdm=100~\gev$, $\mh=150~\gev$ and $\sa=10^{-2}$.  \label{fig:evolgs}}
\end{center}
\end{figure}

To begin with, figure \ref{fig:evolgs} shows the dark matter abundance, $Y=n/s$, as a function of the temperature for different values of $g_s$. The other parameters were chosen as $\mdm=100~\gev$, $\mh=150~\gev$ and $\sa=10^{-2}$. For comparison the equilibrium abundance, $Y_{eq}$, is also shown (solid orange line).  We see that, as it is typical for FIMPs, the dark matter abundance increases as the Universe cools down until the \emph{freeze-in} temperature, $\tfi$, is reached. Below that temperature the dark matter abundance remains constant because the particles in the thermal plasma no longer have enough energy to produce additional dark matter particles. From the figure it can be seen that, as expected, $\tfi\sim \mdm$ independently of the value of $g_s$.  Notice that, for a given value of $g_s$, the behavior of $Y$ changes from $1/T$ to $1/T^3$ as we go down in temperature, in agreement with the discussion in the previous section.  The point at which the transition between these two regimes occurs depends on $g_s$, moving to higher temperatures as $g_s$ is decreased. It is also observed in the figure that  at intermediate and low  temperatures $Y\propto g_s^2$ whereas  $Y\propto g_s^4$ at very high temperatures. Finally, notice that the non-equilibrium condition, $Y\ll Y_{eq}$ for $T>\tfi$, required by the freeze-in mechanism is easily satisfied for all the values of $g_s$ considered in the figure.

\begin{figure}[t]
\begin{center} 
\includegraphics[scale=0.4]{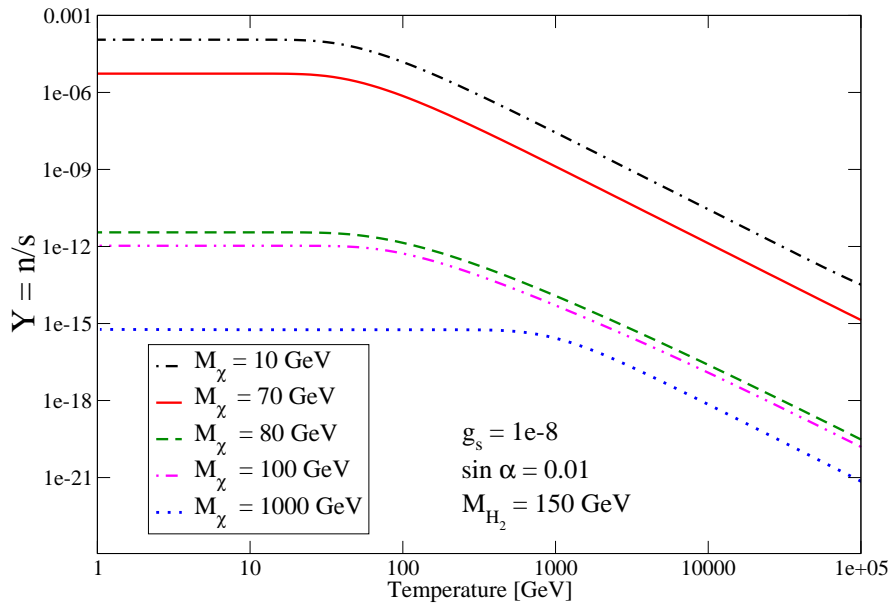}
\caption{\small The dark matter abundance as a function of the temperature for different values of $\mdm$. In this figure  $g_s=10^{-8}$, $\mh=150~\gev$ and $\sa=10^{-2}$.  \label{fig:evolmchi}}
\end{center}
\end{figure}

The dependence of $Y$  on the dark matter mass is illustrated in figure \ref{fig:evolmchi}, which shows $Y$ as a function of the temperature for different values of $\mdm$. In it we set  $g_s=10^{-8}$, $\sa=10^{-2}$, and $\mh=150~\gev$. This value of $\mh$ implies that only for the two smallest values of $\mdm$ ($10~\gev$ and $70~\gev$) $\sv$ can be enhanced thanks to the  $H_2$ resonance --via diagram (c). From the figure we see that   
it makes a huge difference in the value of $Y$ whether this resonant enhancement can take place or not. In fact, models that benefit from the enhancement have an abundance about six orders of magnitude larger than those that do not --see the lines for $\mdm=70~\gev$ and $\mdm=80~\gev$. Note that the freeze-in temperature does depend on $\mdm$, and it is given by $\tfi\sim \mdm$ at large masses (see the bottom line). For masses below $10~\gev$ or so, the abundance becomes independent of the dark matter mass so the line for $\mdm=10~\gev$ (upper one) can actually be interpreted as valid for any $\mdm<10~\gev$. At the other end of the spectrum, we see that the abundance decreases as the dark matter mass in increased --compare $\mdm=100~\gev$ (magenta line) with $\mdm=1~\tev$ (blue dotted-line).

\begin{figure}[t]
\begin{center} 
\includegraphics[scale=0.4]{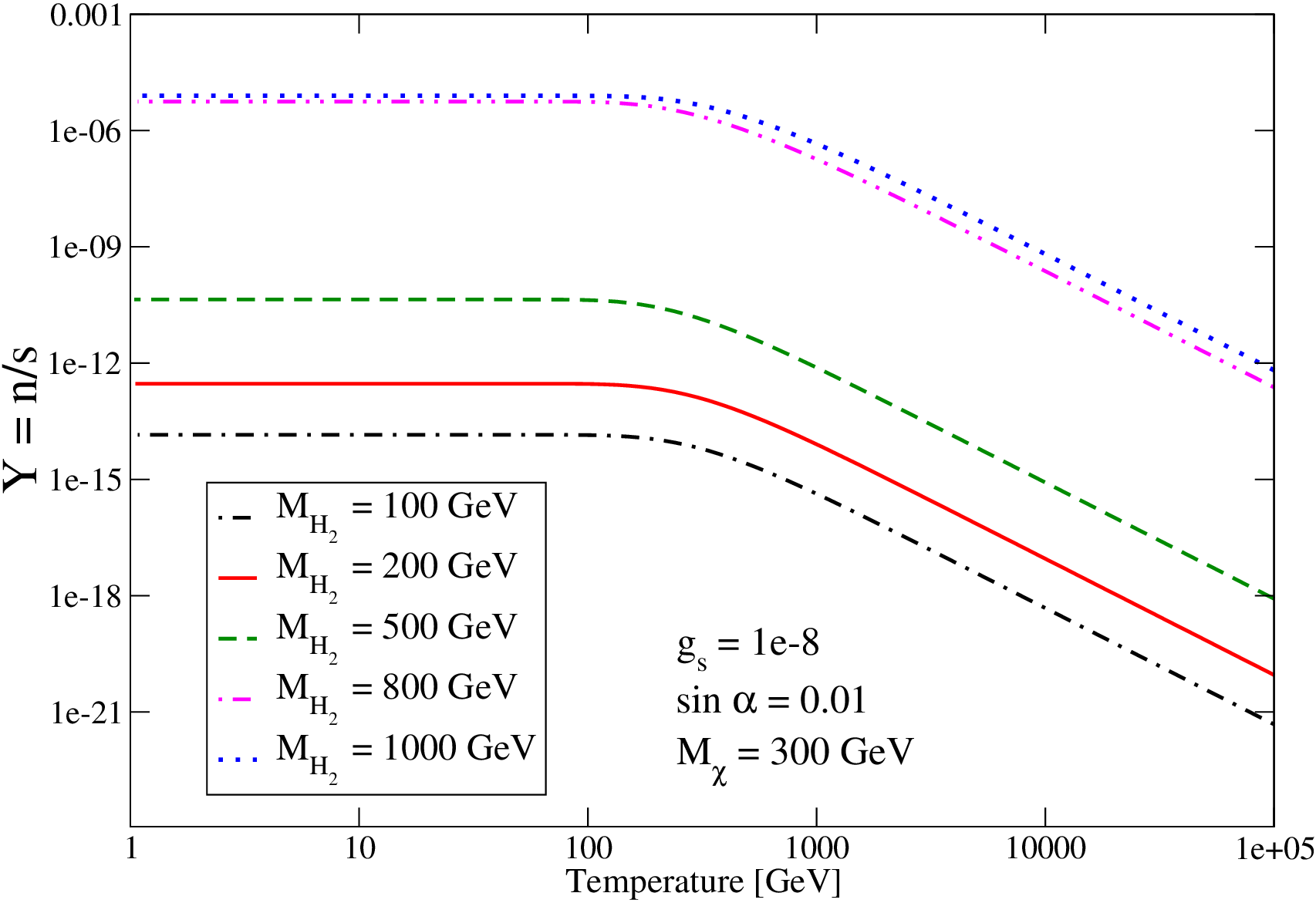}
\caption{\small The dark matter abundance as a function of the temperature for different values of $\mh$. In this figure  $g_s=10^{-8}$, $\mdm=300~\gev$ and $\sa=10^{-2}$.  \label{fig:evolmh2}}
\end{center}
\end{figure}

Figure \ref{fig:evolmh2} displays the dependence of $Y$ with $\mh$. The other parameters were taken as $g_s=10^{-8}$, $\sa=10^{-2}$ and $\mdm=300~\gev$. Again we notice a huge difference in $Y$ between models where dark matter can be produced resonantly (two upper lines) and those where it cannot (three lower lines). As expected, the freeze-in temperature is determined by $\mdm$ and does not depend on $\mh$. We also observed from the figure that $Y$ increases with $\mh$. 

\begin{figure}[t]
\begin{center} 
\includegraphics[scale=0.4]{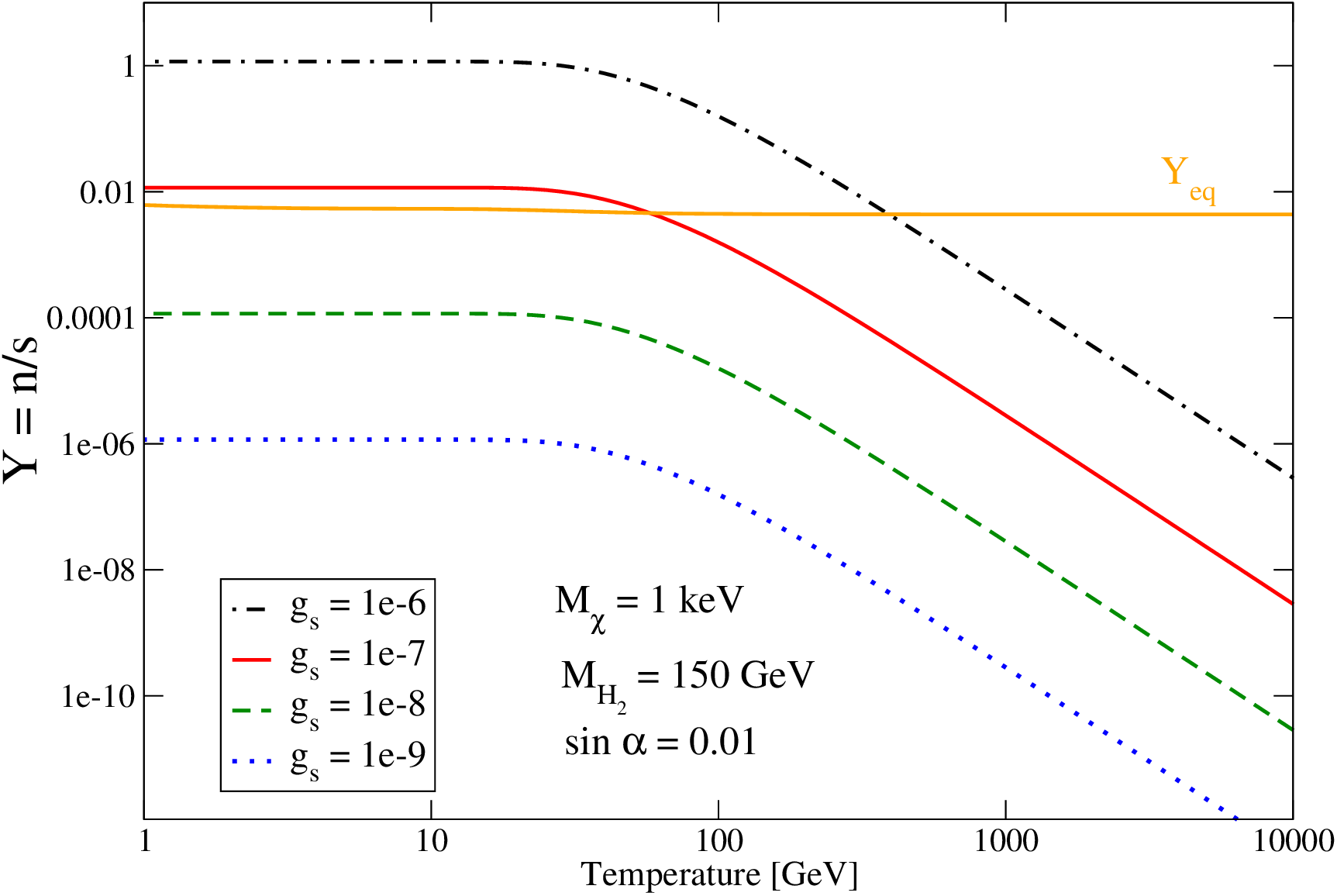}
\caption{\small The dark matter abundance as a function of the temperature for different values of $g_s$. The solid orange line shows $Y_{eq}$. In this figure  $\mdm=1~\kev$, $\mh=150~\gev$ and $\sa=10^{-2}$.  \label{fig:evolgsw}}
\end{center}
\end{figure}

In the previous figures, we have considered dark matter masses in the GeV-TeV range, which is the usual scale associated with WIMPs.  Another interesting possibility is to have a dark matter candidate at the keV scale, so that it behaves as $\emph{warm}$ dark matter. Figure \ref{fig:evolgsw} shows the evolution of the dark matter abundance for $\mdm=1~\kev$, $\mh=150~\gev$,  $\sa=10^{-2}$ and different values of $g_s$.  In this case, the production is always dominated by the resonance and the freeze-in temperature is not determined by $\mdm$, lying instead in the $10$ to $100$ GeV range. Thus, the freeze-in takes place while the dark matter particles are very relativistic. For comparison, the equilibrium abundance, $Y_{eq}$ (solid orange line), is also displayed in the figure. Notice that for $g_s\geq 10^{-7}$, $Y>Y_{eq}$ for $T<\tfi$, contradicting the assumption made in equation (\ref{eq:boltzmann}). For such values of $g_s$, therefore, the FIMP framework does not work and our predictions are not reliable. If, on the other hand, $g_s\lesssim 10^{-8}$, $Y\ll Y_{eq}$ in agreement with our assumptions. As we will see in \ref{sec:vr}, for dark matter masses in the keV range the value of $g_s$ required to obtain the observed relic abundance turns out to be close to $10^{-8}$, so  it is safely within the FIMP region. It might have been otherwise though.

\begin{figure}[t]
\begin{center} 
\includegraphics[scale=0.4]{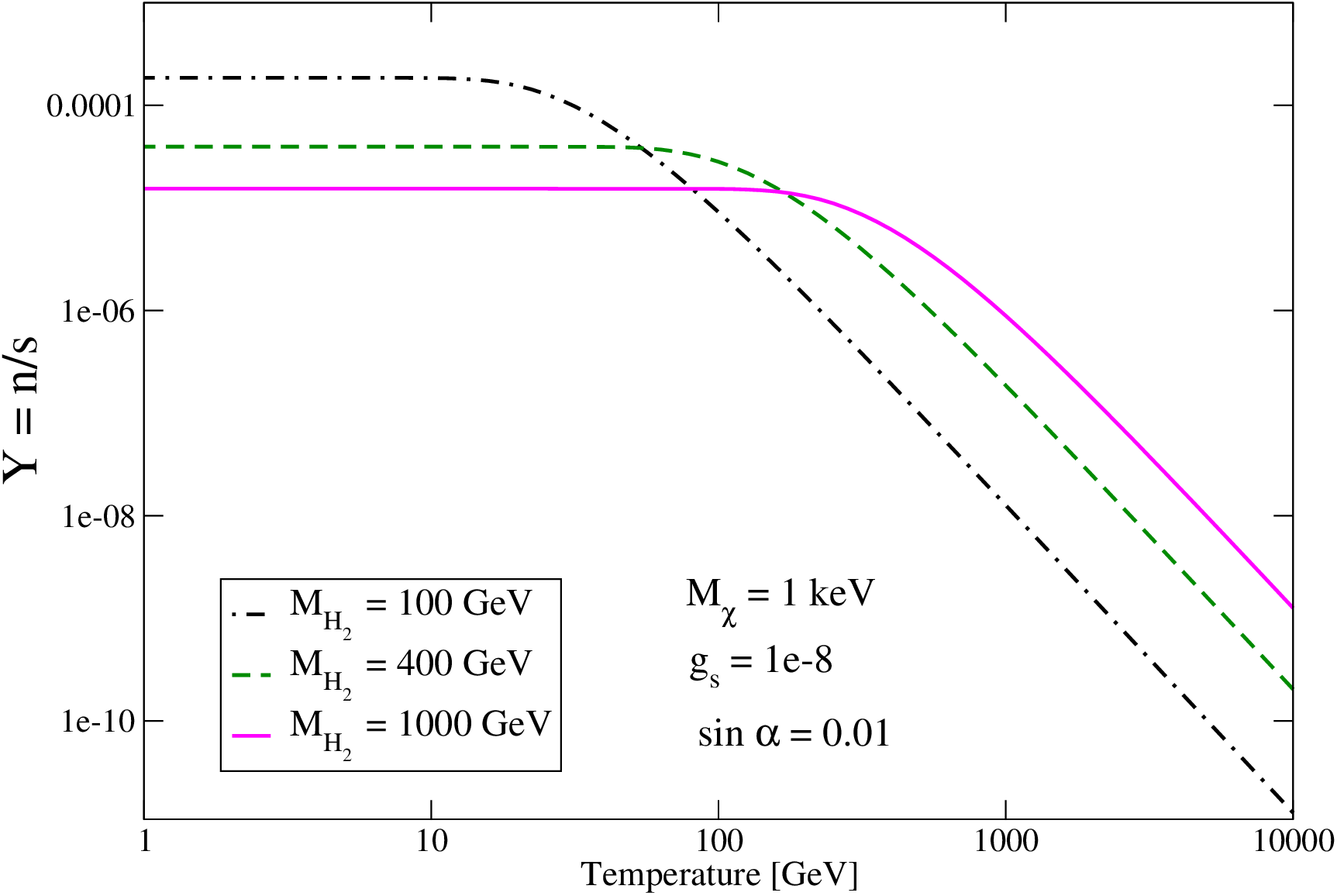}
\caption{\small The dark matter abundance as a function of the temperature for different values of $\mh$. In this figure  $g_s=10^{-8}$, $\mdm=1~\kev$ and $\sa=10^{-2}$.  \label{fig:evolmh2w}}
\end{center}
\end{figure}
In the warm dark matter regime, the only other parameter that affects $Y$ is $\mh$. Figure \ref{fig:evolmh2w} shows the variation of $Y$ with the temperature for different values of $\mh$. $\mdm$, $g_s$ and $\sa$ were set respectively to $1~\kev$, $10^{-8}$, and $10^{-2}$. We see from the figure that the freeze-in temperature increases with $\mh$. The asymptotic value of the abundance, on the contrary, decreases with it.

As we have seen, in the FIMP regime the dark matter abundance depends non-trivially on the temperature and on the parameters of the singlet fermionic model. It goes from $Y\propto 1/T$ at very high temperatures to $Y\propto 1/T^3$ at intermediate ones, reaching the freeze-in temperature and remaining constant afterward. It increases either with $g_s^2$ or $g_s^4$,  it varies with $\mdm$ and $\mh$, and it is very sensitive to the effect of the $H_2$ resonance. In the next section, we  turn our attention to the relic density, the quantity that is actually constrained by observational data.

\subsection{The relic density}
\label{sec:rd}

\begin{figure}[t]
\begin{center} 
\includegraphics[scale=0.4]{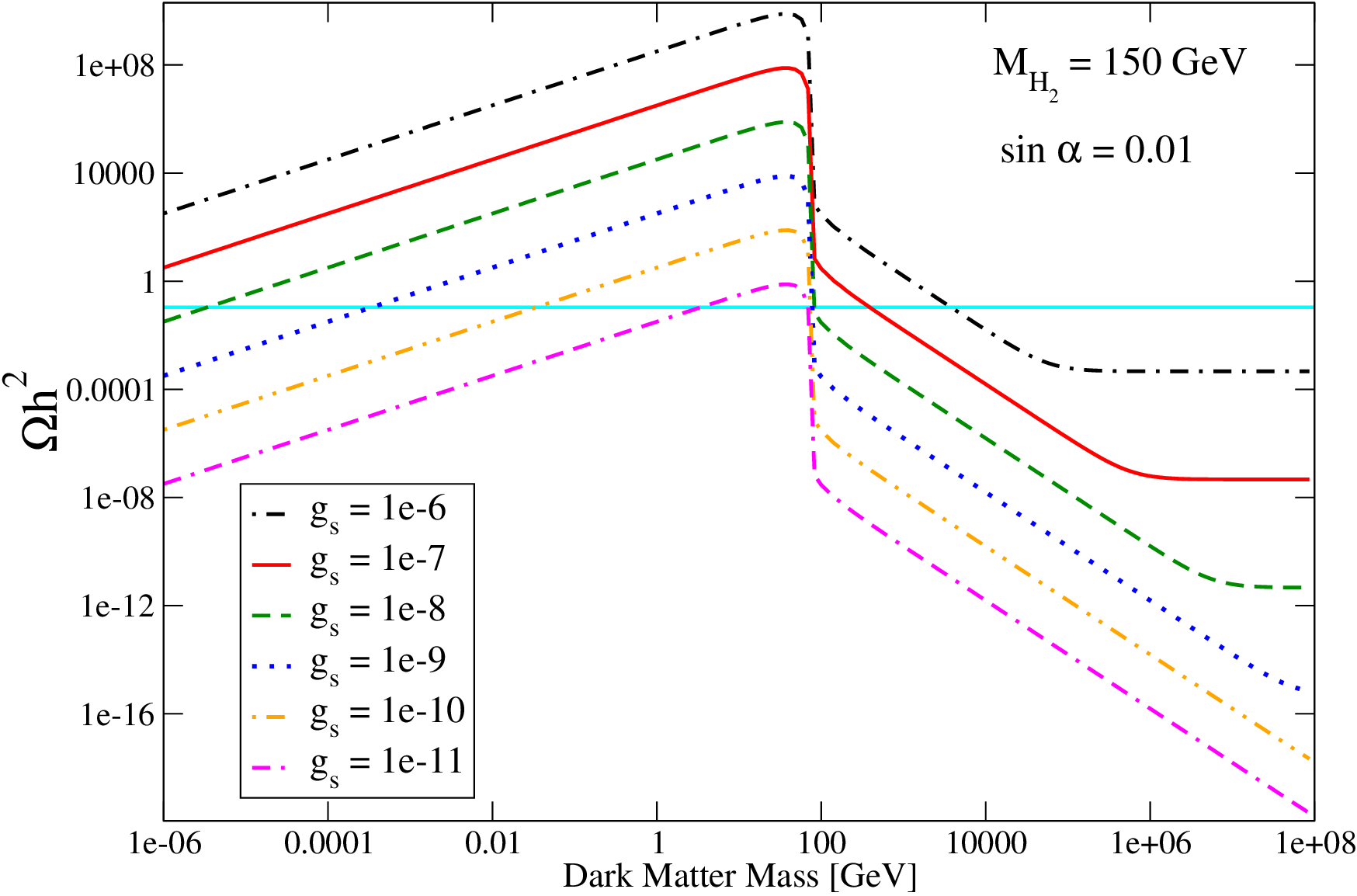}
\caption{\small The relic density  as a function of the dark matter mass for different values of $g_s$.  In this figure  $\mh=150~\gev$ and $\sa=10^{-2}$. The horizontal  band corresponds to the observed value of the dark matter density. \label{fig:rdgs}}
\end{center}
\end{figure}

The present relic density of dark matter is proportional to $\mdm$ and to the asymptotic value of $Y$, see equation (\ref{eq:rd}). Figure \ref{fig:rdgs} shows the relic density as a function of the dark matter mass for different values of $g_s$. The other parameters were taken as $\mh=150~\gev$ and $\sa=10^{-2}$. For the dark matter mass we consider a very wide range starting at the $\kev$, passing through the weak scale and going up to $10^{8}~\gev$. The lines correspond to values of $g_s$ varying by one order of magnitude from $10^{-6}$ (top line) to $10^{-11}$ (bottom line). Depending on the dark matter mass, the behavior of  the relic density can be divided into four regions: below the resonance, the resonance, above the resonance, and the ultra-heavy regime.  The region below the resonance, which includes the warm dark matter region,  is characterized by a relic density that increases linearly with the dark matter mass. At the resonance, $\mdm\sim \mh/2$, the relic density varies by several orders of magnitude within a very small range of $\mdm$. Above the resonance the relic density decreases with $\mdm^2$. Finally, if $g_s$ is large enough, we also observe a region where the relic density becomes independent of the dark matter mass. We call that region the ultra-heavy regime, as it opens up for masses above $10^5$ GeV. Notice that $\oh\propto g_s^2$ except in the ultra-heavy regime, where  $\oh\propto g_s^4$. The reason for this behavior is that in this regime dark matter is dominantly produced via $H_2H_2$-annihilation --see diagram (d) in figure \ref{fig:diag1}-- whose cross section indeed goes as $g_s^4$. The horizontal cyan band in this figure corresponds to the observed value of the dark matter density \cite{Hinshaw:2012aka,Ade:2013lta}. We see that compatibility with current data requires $g_s\sim 10^{-8}$ for $\kev$ masses whereas values above $10^{-6}$ would be  needed in the ultra-heavy regime. 

\begin{figure}[t]
\begin{center} 
\includegraphics[scale=0.4]{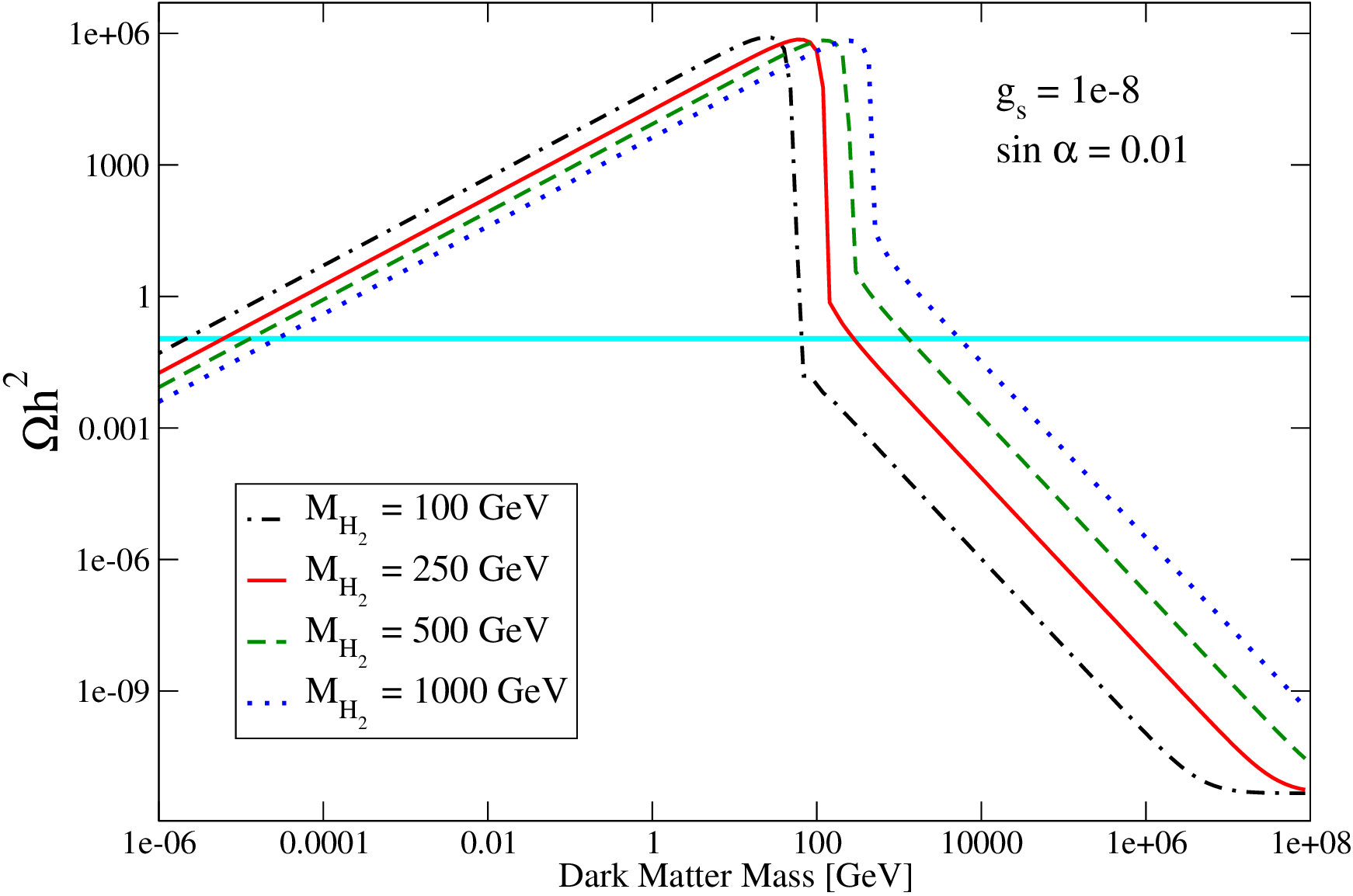}
\caption{\small The relic density  as a function of the dark matter mass for different values of $\mh$. In this figure  $g_s=10^{-8}$ and $\sa=10^{-2}$. The horizontal  band corresponds to the observed value of the dark matter density.  \label{fig:rdmh2}}
\end{center}
\end{figure}

Next, we illustrate, in figure \ref{fig:rdmh2}, the effect of $\mh$ on the relic density for fixed values of $g_s$ and $\sa$ --respectively $10^{-8}$ and $10^{-2}$. As expected, $\mh$ sets the value of the dark matter mass where the resonance is encountered,  and therefore it determines the transition between the below the resonance and the above the resonance regions. In the former the relic density decreases when $\mh$ is increased whereas the opposite behavior is observed above the resonance. For this value of $g_s$, the relic density agrees with the observations in two different mass regimes. One corresponds to very light masses, between $1~\kev$ and $100~\kev$. The other one  lies around the weak scale, with masses between $100~\gev$ and $10~\tev$.  

\begin{figure}[t]
\begin{center} 
\includegraphics[scale=0.4]{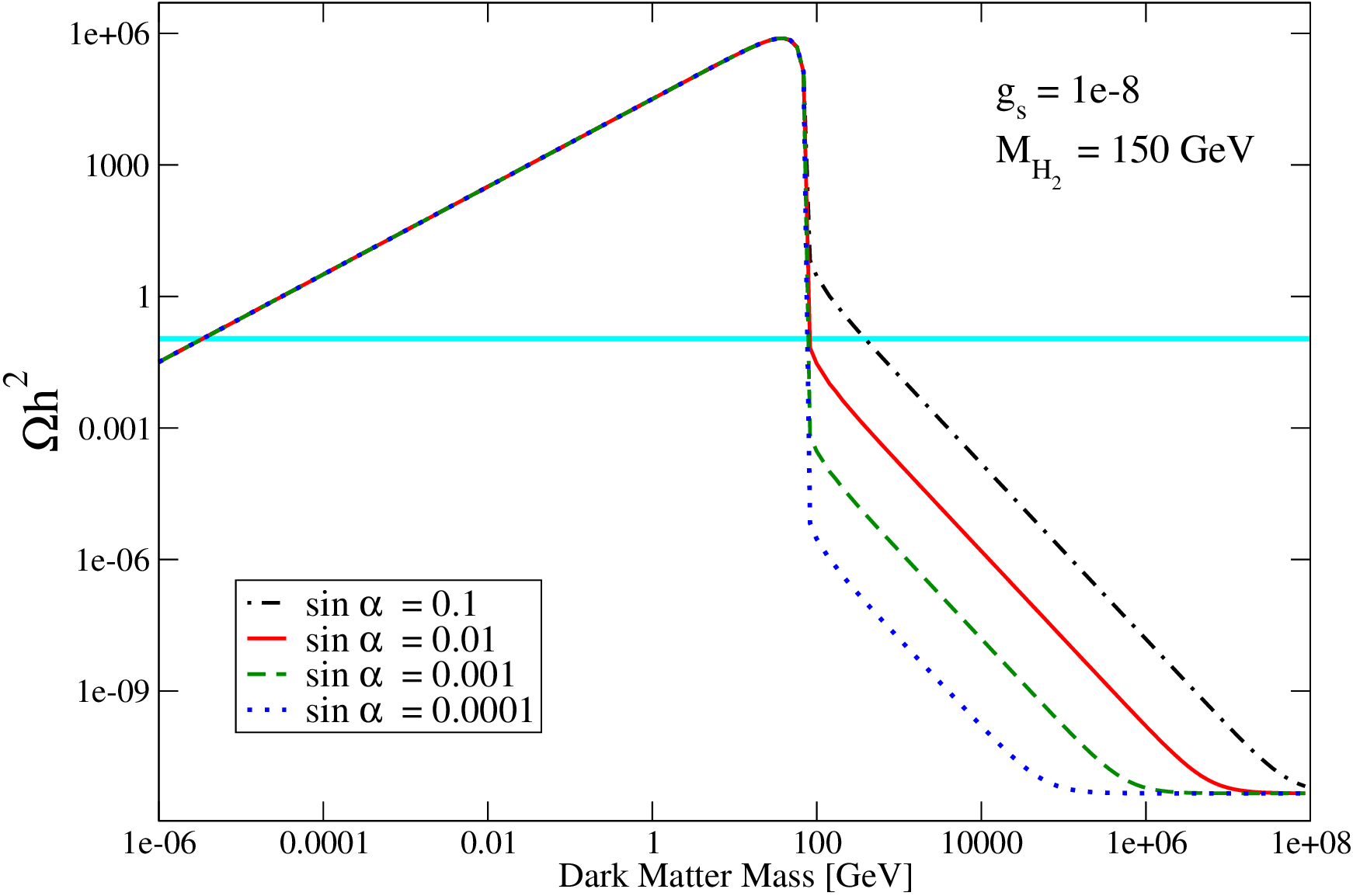}
\caption{\small The relic density  as a function of the dark matter mass for different values of $\sa$. In this figure  $g_s=10^{-8}$ and $\mh=150~\gev$. The horizontal  band corresponds to the observed value of the dark matter density. \label{fig:rdsa}}
\end{center}
\end{figure}

The variation of the relic density with $\sa$ is shown in figure \ref{fig:rdsa}. In contrast with $g_s$ and $\mh$,  only the resonance and the above the resonance regions are affected by $\sa$. Below the resonance and in the ultra-heavy regime the relic density is insensitive to this parameter. For warm dark matter, in particular, the relic density does not depend on $\sa$. Notice from the figure that $\sa$ also affects the value of the dark matter mass  where the ultra-heavy regime starts. This is to be expected as a smaller value of $\sa$ allows the contribution from the (d) diagram in figure \ref{fig:diag1}  to dominate at lower masses.

\begin{figure}[tb]
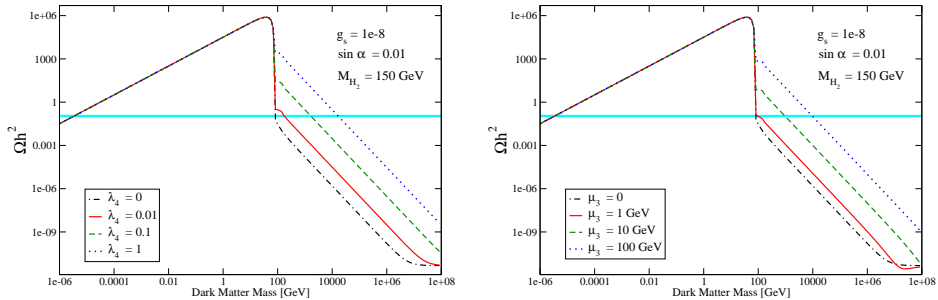

\begin{center} 
\begin{tabular}{cc}
\includegraphics[scale=0.2]{rdla} & \includegraphics[scale=0.2]{rdmu3}
\end{tabular}
\caption{\small The relic density  as a function of the dark matter mass for different values of $\lambda_4$ (left) and $\mu_3$ (right).  In this figure  $g_s=10^{-8}$, $\mh=150~\gev$, and $\sa=10^{-2}$. The horizontal  band corresponds to the observed value of the dark matter density. \label{fig:rdother}}
\end{center}
\end{figure}

In our previous figures, we have always used what we call our default setup: $\lambda_4=0$ and $\mu_3=0$. If these couplings are non-zero new processes contribute to the production of dark matter. $\lambda_4$, for instance, allows for dark matter production via $H_1H_2$ annihilation (see diagram (e) in figure \ref{fig:diag1}) whereas $\mu_3$ contributes to dark matter production via $H_2H_2$ annihilation (diagram (f) in figure \ref{fig:diag1}). They are expected, therefore, to increase the dark matter relic density. Figure \ref{fig:rdother} shows the relic density for different values of $\lambda_4$ (left panel) and of $\mu_3$ (right panel). As was the case with $\sa$, these two parameters only affect the relic density in the region above the resonance. 

We have thus analyzed the dependence of the dark matter relic density  with respect to all the free parameters of the singlet fermionic model. Depending on the dark matter mass, four different regions can be recognized: Below the resonance, where the relic density increases linearly with $\mdm$; the resonance, in which the relic density changes by several orders of magnitude within a small range of $\mdm$; above the resonance, where $\oh$ decreases as $1/\mdm^2$; and the ultra-heavy regime in which $\oh$ becomes independent of $\mdm$. Both $g_s$ and $\mh$ were shown to affect the value of the  relic density. A major simplification is the fact that $\sa$, $\lambda_4$ and $\mu_3$ only modify the relic density in a restricted area of the parameter space --what we call the above the resonance region. In the next section, we will use the value of the relic density to obtain the viable regions of the singlet fermionic model in the FIMP regime. That is, the regions of the parameter space that are consistent with the dark matter constraint.

\subsection{The viable regions}
\label{sec:vr}
\begin{figure}[t]
\begin{center} 
\includegraphics[scale=0.4]{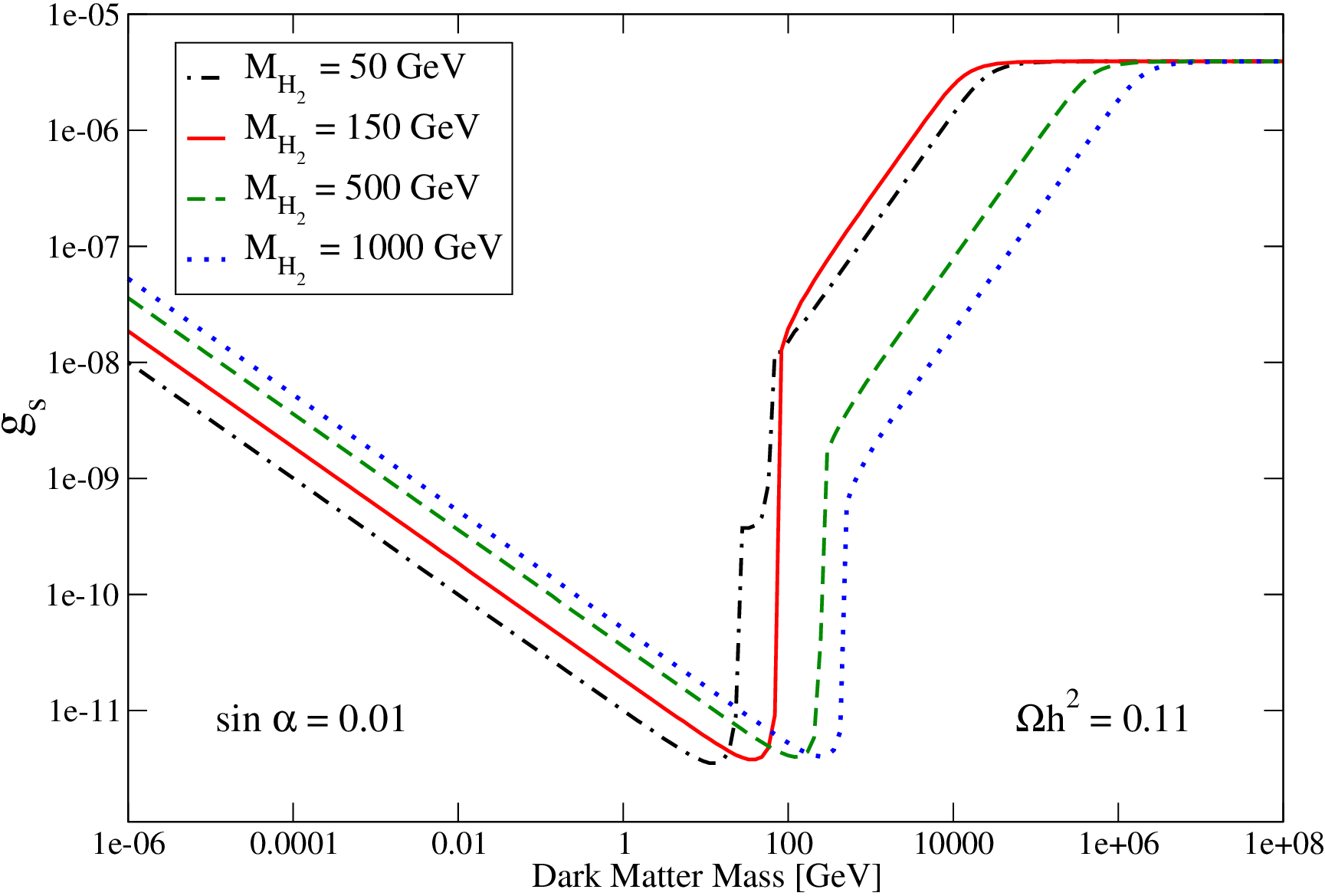}
\caption{\small The viable regions in the plane ($g_s$, $\mdm$)  for different values of $\mh$. In this figure $\sa=10^{-2}$. Along the lines $\oh=0.11$. \label{fig:levelmh2}}
\end{center}
\end{figure}

The most important parameters of the singlet fermionic model are the dark matter mass, $\mdm$, and the dark matter coupling, $g_s$. It makes sense, therefore, to project the multidimensional viable regions onto the plane ($\mdm$, $g_s$). That is precisely what we do in this section: we display, in the plane ($\mdm$,$g_s$), the lines that are consistent with the observed value of the dark matter density, $\oh=0.11$, for given values of the remaining parameters of the model: $\sa$, $\mh$, $\lambda_4$ and $\mu_3$. In our calculations we also require that all viable points satisfy the condition $Y\ll Y_{eq}$ for  $T>\tfi$. This condition is necessary to ensure that we are indeed dealing with a freeze-in scenario and that equation (\ref{eq:boltzmann}) can be used.

Figure \ref{fig:levelmh2} shows the viable regions for different values of  $\mh$. In it,  $\sa$ was set to $10^{-2}$ and values of $\mh$ up to $1~\tev$ were considered. For $\kev$ dark matter masses we see that $g_s$ should be of order $10^{-8}$, with a slight dependence on the value of $\mh$.  At even smaller masses, the coupling required would be larger and, as illustrated in figure \ref{fig:evolgsw}, we would be violating the conditions of the freeze-in mechanism. The $\kev$ scale seems, therefore, to be special in this model because it is the smallest dark matter mass that is compatible with the FIMP framework. As the dark matter mass increases, the value of $g_s$ required decreases reaching a minimum of $\sim 3\times 10^{-12}$ right before the resonance. At the resonance $g_s$  increases significantly and it continues to increase, though not as quickly, above the resonance. Then at masses of order $10^5$-$10^{6}$ GeV, $g_s$ reaches its maximum value ($\sim 4\times 10^{-6}$) and remains constant as the dark matter mass is further increased. The value of $\mh$  slightly modifies all the viable regions except at very large dark matter masses.  The FIMP framework is thus able to explain the observed dark matter density over a huge range of masses, from the $\kev$ scale to  well-above the electroweak scale. 

\begin{figure}[t]
\begin{center} 
\includegraphics[scale=0.4]{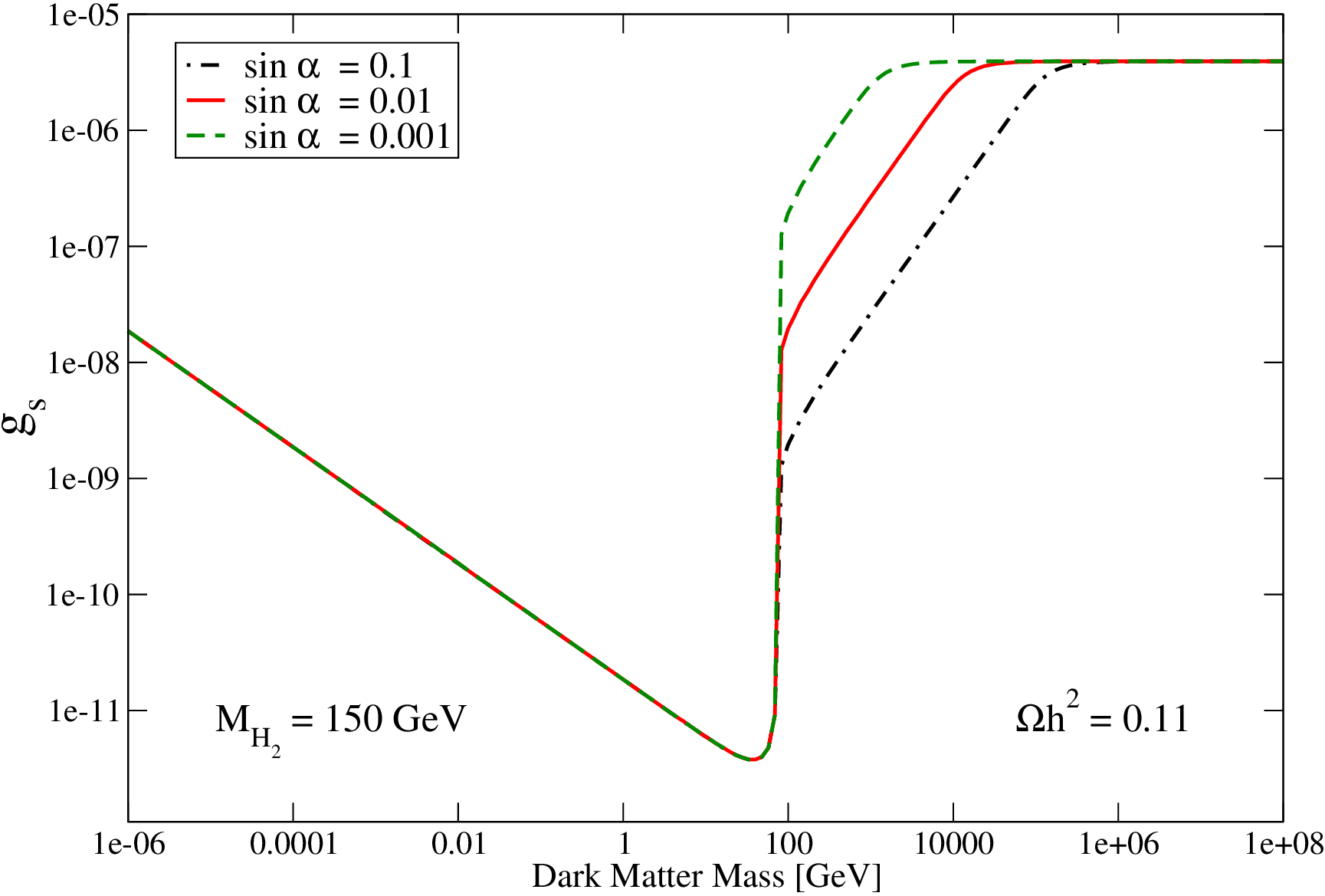}
\caption{\small The viable regions in the plane ($g_s$, $\mdm$)  for different values of $\sa$. In this figure  $\mh=150~\gev$.  Along the lines $\oh=0.11$. \label{fig:levelsa}}
\end{center}
\end{figure}

In view of this result, one may wonder whether there is an upper limit to the mass of the dark matter particle within this framework. It is known, for example, that in the WIMP regime unitarity imposes an upper bound on the dark matter mass of about $120~\tev$\cite{Griest:1989wd}. Clearly, such a limit does not apply to the freeze-in scenario. There is, however, a different kind of limit that does apply to FIMPs and it is set by the reheating temperature of the Universe $T_{RH}$ --a  currently unknown cosmological parameter. The standard freeze-in process implicitly assumes, via the initial condition for the Boltzmann equation, that $\mdm<T_{RH}$.  If that is not the case, a slight variation of the freeze-in scenario is obtained, as studied in \cite{Yaguna:2011ei} for the singlet scalar model. In this paper we have limited ourselves to the standard freeze-in framework and have assumed that $T_{RH}>\mdm$ in all cases.

Figure \ref{fig:levelsa} shows the viable lines  for $\mh=150~\gev$ and different values of $\sa$. As we have noted in the previous section, only the region above the resonance is affected by $\sa$. For a given $\mdm$, a larger value of $\sa$ implies a proportionally smaller value of $g_s$. Notice that for $\sa= 10^{-3}$ the ultra-heavy regime, where $g_s$ remains constant as the dark matter mass in increased, actually starts below $5~\tev$. Thus, by reducing the value of $\sa$ one can bring the ultra-heavy regime close to the electroweak scale.

\begin{figure}[tb]
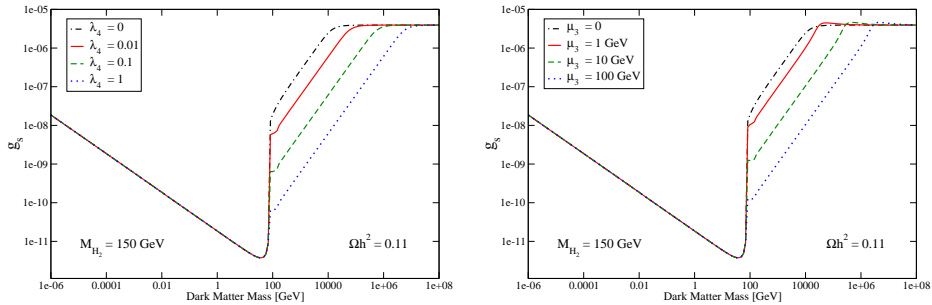

\begin{center} 
\begin{tabular}{cc}
\includegraphics[scale=0.2]{levella} & \includegraphics[scale=0.2]{levelmu3}
\end{tabular}
\caption{\small The viable regions in the plane ($g_s$, $\mdm$)  for different values of $\lambda_4$ (left) and $\mu_3$ (right). In this figure $\mh=150~\gev$ and $\sa=10^{-2}$. Along the lines $\oh=0.11$. \label{fig:levelother}}
\end{center}
\end{figure}

Finally, figure \ref{fig:levelother} shows the viable regions for different values  of $\lambda_4$ (left panel) and $\mu_3$ (right panel). Only the region above the resonance is affected by these variables, and in the expected way --see previous section. Notice that a non-zero value for these parameters increases the value of  $\mdm$ where the ultra-heavy mass regime starts.

\section{Discussion}
\label{sec:dis}
As we had  demonstrated in the previous section, the freeze-in mechanism can indeed explain the observed dark matter density within the singlet fermionic model. It is important to emphasize that this alternative scenario for dark matter production  is as simple and predictive as the WIMP framework. Besides the given particle physics model, they assume only the Standard Cosmological Model --e.g. a radiation dominated Universe during  dark matter production or freeze-out. From the dark matter point of view, henceforth, they  should be treated on the same footing. There is no objective reason to prefer the WIMP solution studied previously over the FIMP one considered in this paper.  

Their implications, on the other hand, are quite different. Whereas the WIMP regime of the singlet fermionic model  is strongly constrained by direct detection experiments --see e.g.\cite{Esch:2013rta}--, the FIMP regime we have analyzed is, in virtue of the small values of $g_s$ involved, essentially insensitive to them. Not even future experiments are expected to directly probe the viable regions in a meaningful way. But if a credible signal of dark matter is observed in direct or indirect dark matter experiments, the FIMP regime of the singlet fermionic model can be immediately excluded  as the explanation of dark matter. In this sense, it is also a testable and falsifiable scenario. 

We already pointed out that  the $\kev$ scale seems to play a special role within this model because it is the smallest mass scale that can be compatible with the freeze-in mechanism.  Smaller masses would require larger values of $g_s$, which would in turn bring the dark matter particle into equilibrium, invalidating the fundamental assumption of the freeze-in. Remarkably, the $\kev$  scale is also the mass scale associated with warm dark matter. Thus, it is a non-trivial fact that both cold and warm dark matter can be obtained via freeze-in in the singlet fermionic model. This result also shows that this mechanism not only works for massive particles, as originally put forward in \cite{Hall:2009bx}. Even though many models of warm dark matter have been proposed in recent years \cite{Barry:2011fp,Ma:2012if,Nemevsek:2012cd,King:2012wg,
Merle:2013wta}, they do not make use of the freeze-in mechanism, often relying instead on thermal overproduction followed by entropy dilution. To our knowledge, the singlet fermionic model we have considered is the first application of the freeze-in mechanism to the direct production of a  warm dark matter particle.

Because the  freeze-in process requires a dark matter particle that is  singlet under the SM gauge group, the simplest realizations of this scenario contain a singlet scalar or a singlet fermion as the dark matter candidate. The former possibility had been previously addressed in \cite{Yaguna:2011qn,Yaguna:2011ei} while the latter has been examined in this paper. Surprisingly, there are substantial differences between the results obtained in both realizations. What we call the ultra-heavy regime in the singlet fermionic model, for instance, actually starts right above the resonance in the singlet scalar case, and it features much smaller values of the coupling. The resonance is determined by $H_2$ in the  fermionic case but  by the Higgs boson   in the scalar case. As a result, the variation of the relic density through the resonance is quite different in both models. At the end, these deviations can be traced back to the fact that  the fermionic model, being slightly more complex, offers additional possibilities. In contrast to the scalar case, for example, the production of dark matter in the singlet fermionic model  is determined by different processes depending on the temperature and the parameters of the model. In addition, the singlet scalar model cannot easily accommodate warm dark matter because the singlet mass receives a contribution after electroweak symmetry breaking that, being proportional to  $\sqrt\lambda v$, is always  much larger than a $\kev$. Hence, the singlet fermionic model that we have studied provides an alternative and phenomenologically rich illustration of the freeze-in paradigm.

\section{Conclusions}
\label{sec:con}
We have studied the FIMP realization of the singlet fermionic model of dark matter. This model, a minimal extension of the SM that can also account for the dark matter, had been studied before but only as a WIMP model, where the dark matter density is the result of a freeze-out. In this paper, we have for the first time studied the possibility of explaining the dark matter via the freeze-in mechanism instead. Within that framework, we analyzed the dependence of the dark matter abundance and the relic density  on the different parameters of the model, and we obtained the regions that are compatible with the dark matter constraint.  These viable regions, as expected, are quite different from those obtained within the WIMP framework, featuring much smaller values of the $g_s$ coupling, between $10^{-12}$ and $10^{-6}$.  As we have demonstrated, the observed dark matter density can be obtained for a wide range of masses, starting  from  super-heavy dark matter,  passing through the $\tev$ scale,  and going all the way down to the $\kev$ scale.  Hence, in the freeze-in realization of the singlet fermionic model, it is possible to obtain both cold and  warm  dark matter.

\section*{Acknowledgments}
This work  is partially supported by the ``Helmholtz Alliance for Astroparticle Phyics HAP''
 funded by the Initiative and Networking Fund of the Helmholtz Association. 
\bibliographystyle{hunsrt}
\bibliography{darkmatter}
\end{document}